\PassOptionsToPackage{unicode}{hyperref}
\PassOptionsToPackage{hyphens}{url}
\PassOptionsToPackage{dvipsnames,svgnames,x11names}{xcolor}
\documentclass[]{article}
\usepackage[a4paper,margin=1in]{geometry}
\usepackage{amsmath,amssymb}
\usepackage{lmodern}
\usepackage{iftex}
\usepackage[T1]{fontenc}
\usepackage[utf8]{inputenc}
\usepackage{textcomp} 
\IfFileExists{upquote.sty}{\usepackage{upquote}}{}
\IfFileExists{microtype.sty}{
  \usepackage[]{microtype}
  \UseMicrotypeSet[protrusion]{basicmath} 
}{}
\makeatletter
\@ifundefined{KOMAClassName}{
  \IfFileExists{parskip.sty}{%
    \usepackage{parskip}
  }{
    \setlength{\parindent}{0pt}
    \setlength{\parskip}{6pt plus 2pt minus 1pt}}
}{
  \KOMAoptions{parskip=half}}
\makeatother
\usepackage{xcolor}
\usepackage{graphicx}
\makeatletter
\def\maxwidth{\ifdim\Gin@nat@width>\linewidth\linewidth\else\Gin@nat@width\fi}
\def\maxheight{\ifdim\Gin@nat@height>\textheight\textheight\else\Gin@nat@height\fi}
\makeatother
\setkeys{Gin}{width=\maxwidth,height=\maxheight,keepaspectratio}
\makeatletter
\def\fps@figure{htbp}
\makeatother
\providecommand{\tightlist}{%
  \setlength{\itemsep}{0pt}\setlength{\parskip}{0pt}}
\NewDocumentCommand\citeproctext{}{}
\NewDocumentCommand\citeproc{mm}{%
  \begingroup\def\citeproctext{#2}\cite{#1}\endgroup}
\makeatletter
 \let\@cite@ofmt\@firstofone
 \def\@biblabel#1{}
 \def\@cite#1#2{{#1\if@tempswa , #2\fi}}
\makeatother
\newlength{\cslhangindent}
\newlength{\csllabelwidth}
\newenvironment{CSLReferences}[2] 
 {\begin{list}{}{%
  \setlength{\itemindent}{0pt}
  \setlength{\leftmargin}{0pt}
  \setlength{\parsep}{0pt}
  \ifodd #1
   \setlength{\leftmargin}{\cslhangindent}
   \setlength{\itemindent}{-1\cslhangindent}
  \fi
  \setlength{\itemsep}{#2\baselineskip}}}
 {\end{list}}
\usepackage{calc}

\ifLuaTeX
\usepackage[bidi=basic]{babel}
\else
\usepackage[bidi=default]{babel}
\fi
\babelprovide[main,import]{american}

\def\languageshorthands#1{}
\ifLuaTeX
  \usepackage{selnolig}  
\fi
\IfFileExists{bookmark.sty}{\usepackage{bookmark}}{\usepackage{hyperref}}
\IfFileExists{xurl.sty}{\usepackage{xurl}}{} 
\hypersetup{
  pdftitle={Magrathea v2: A planetary interior modeling platform in
C++},
  pdfauthor={David R. Rice, Chenliang Huang, Robert Royer, Mangesh
Daspute, Krishang Mittal, Baptiste Journaux, Jason H. Steffen, Allona
Vazan},
  pdflang={en-US},
  colorlinks=true,
  linkcolor={Maroon},
  filecolor={Maroon},
  citecolor={Blue},
  urlcolor={Blue},
  pdfcreator={LaTeX via pandoc}}

\title{Magrathea v2: A planetary interior modeling platform in C++}

\definecolor{c53baa1}{RGB}{83,186,161}
\definecolor{c202826}{RGB}{32,40,38}

\usepackage[affil-it]{authblk}
\usepackage{orcidlink}
\author[1%
  \ensuremath\mathparagraph]{David R. Rice%
    \,\orcidlink{0000-0001-6009-8685}\,%
    }
\author[2%
  ]{Chenliang Huang%
    \,\orcidlink{0000-0001-9446-6853}\,%
    }
\author[3%
  ]{Robert Royer%
    \,\orcidlink{0000-0002-0900-0192}\,%
    }
\author[4%
  ]{Mangesh Daspute%
    }
\author[5%
  ]{Krishang Mittal%
    }
\author[6%
  ]{Baptiste Journaux%
    \,\orcidlink{0000-0002-0957-3177}\,%
    }
\author[3%
  ]{Jason H. Steffen%
    \,\orcidlink{0000-0003-2202-3847}\,%
    }
\author[1%
  ]{Allona Vazan%
    \,\orcidlink{0000-0001-9504-3174}\,%
    }

\affil[1]{Astrophysics Research Center (ARCO), Department of Natural
Sciences, The Open University of Israel, Raanana 4353701, Israel%
  }
\affil[2]{Shanghai Astronomical Observatory, Chinese Academy of
Sciences, Shanghai 200030, People's Republic of China%
  }
\affil[3]{Department of Physics and Astronomy, University of Nevada, Las
Vegas, 4505 South Maryland Parkway, Las Vegas, NV 89154, USA%
  }
\affil[4]{Department of Physics, Ariel University, Ariel 40700, Israel%
  }
\affil[5]{Department of Astronomy, University of Wisconsin--Madison, 475
N. Charter Street, Madison, WI 53706, USA%
  }
\affil[6]{Department of Earth and Space Science, University of
Washington, Seattle, WA 98195, USA%
  }
\affil[$\mathparagraph$]{Corresponding author: davidr@post.openu.ac.il%
}
\date{Published: 15 September 2026, Journal of Open Source Software}

\begin{document}
\maketitle

\section{Summary}\label{summary}

Magrathea is an open-source C++ code for modeling the internal structure
of differentiated planets. The initial release, Huang et al.
(\citeproc{ref-Huang:2022}{2022}), introduced the base solver, a modular
framework for defining equations of state (EOS) within phase diagrams
for each layer, and a plan for expanding the code. Many of those updates
are now implemented. Magrathea v2 is a more versatile platform that
supports a wider range of compositions, adds new tools for composition
retrieval, and makes it easier for users to adapt the code to their own
models.

\section{Statement of need}\label{statement-of-need}

Constraining a planet's composition is essential for understanding its
formation and evolution. Observations of mass and radius alone are not
sufficient, since many different interiors can yield the same bulk
density. Interior structure solvers are therefore essential tools for
constraining possible compositions. With observational programs
routinely measuring the densities of small to large planets, researchers
require codes with models that are transparent and flexible which can
adapt as our understanding of planet mineralogy changes.

Magrathea\footnote{Magrathea can be accessed at \href{https://github.com/Huang-CL/Magrathea}{https://github.com/Huang-CL/Magrathea}} is designed as such a platform. Rather than enforcing a fixed
planet model, Magrathea provides a framework in which users can define
their own phase diagrams, equations of state (EOS), and thermal
profiles. The goal is not one preferred interior model, but a fast and
readable code base for building and testing many of them. With the
continued expansion of the physics and usability in Version 2.0,
Magrathea helps the community keep up with the growing precision of
exoplanet observations and experimental constraints on planetary
materials.

\section{State of the field}\label{state-of-the-field}

The field includes several open tools for planet interior structure
modeling. We point readers to the large tables of tools in Acuña et al.
(\citeproc{ref-Acuna:2025}{2025}) and Baumeister et al.
(\citeproc{ref-Baumeister:2025}{2025}) for a broader summary, and
highlight four examples here to illustrate the range of current
approaches. GASTLI (\citeproc{ref-Acuna:2025}{Acuña et al., 2025})
focuses on volatile-rich planets and coupled interior--atmosphere
models. ExoPlex (\citeproc{ref-Unterborn:2023}{Unterborn et al., 2023})
emphasizes the mineralogy of predominantly rocky planets. ExoInt
(\citeproc{ref-Wang:2019}{Wang et al., 2019}) adds modules that use
stellar elemental abundances and devolatilization relationships to
constrain planetary bulk composition. PALEOS
(\citeproc{ref-Attia:2026}{Attia et al., 2026}) provides multi-phase,
tabulated equations of state for rapid interpolation in planetary
structure and evolution models.

Within this landscape, Magrathea emphasizes flexibility in constructing
differentiated planet models and provides a growing set of run modes for
different use cases. Users have explicit control over phase diagrams and
EOS choices in each differentiated layer allowing for diverse planet
models from sub-Earth to Neptune-mass planets. Magrathea allows users to
swap mineral physics assumptions quickly, run forward models fast enough
for large sweeps, and test how interior assumptions propagate into
inferred compositions. The combination of modularity and speed from the
C++ implementation is the main reason we seek to build on Magrathea.

\section{Software design}\label{software-design}

The core solver of Magrathea is a one-dimensional, spherically symmetric
integrator of the equations of hydrostatic equilibrium, mass continuity,
temperature gradient, and equation of state. For a user-defined planet
consisting of up to four differentiated layers, the code integrates
inward and outward solutions using a shooting-to-fitting-point method
with adaptive Runge--Kutta--Fehlberg stepping. The solver returns the
radius of the planet, the radii of each compositional boundary, and
profiles of pressure, temperature, density, and phase as functions of
enclosed mass. Solving one planet takes about one second for most
configurations.

A key design choice is modularity. A large variety of EOS forms are
supported in \texttt{EOS.cpp}, including Birch--Murnaghan, Vinet,
Holzapfel, Keane, ideal and van der Waals gases, Debye or Einstein
thermal terms, and tabulated EOS. Parameters for each material are
defined in a library of more than 70 EOS in \texttt{EOSlist.cpp}. Phase
diagrams for each layer define which material is used at a given
pressure--temperature condition in \texttt{phase.cpp}. Alternative phase
diagrams are also stored in a library and can be selected at run time.

Magrathea separates the solver from the model. The code offers nine run
modes through human-readable \texttt{.cfg} files, including a full
four-layer solver, isothermal and two-layer modes, bulk runs,
two/three-layer composition finder, on-the-fly EOS modification modes,
and an MCMC retrieval mode. This range of models and use cases makes
Magrathea not just a solver but a platform for exploring interior
models.

\section{Major updates in this
version}\label{major-updates-in-this-version}

Since the initial release (\citeproc{ref-Huang:2022}{Huang et al.,
2022}), Magrathea has undergone expansions in physics, solvers, and
usability.

\textbf{New physical models and materials}

\begin{figure}
\centering
\includegraphics[keepaspectratio]{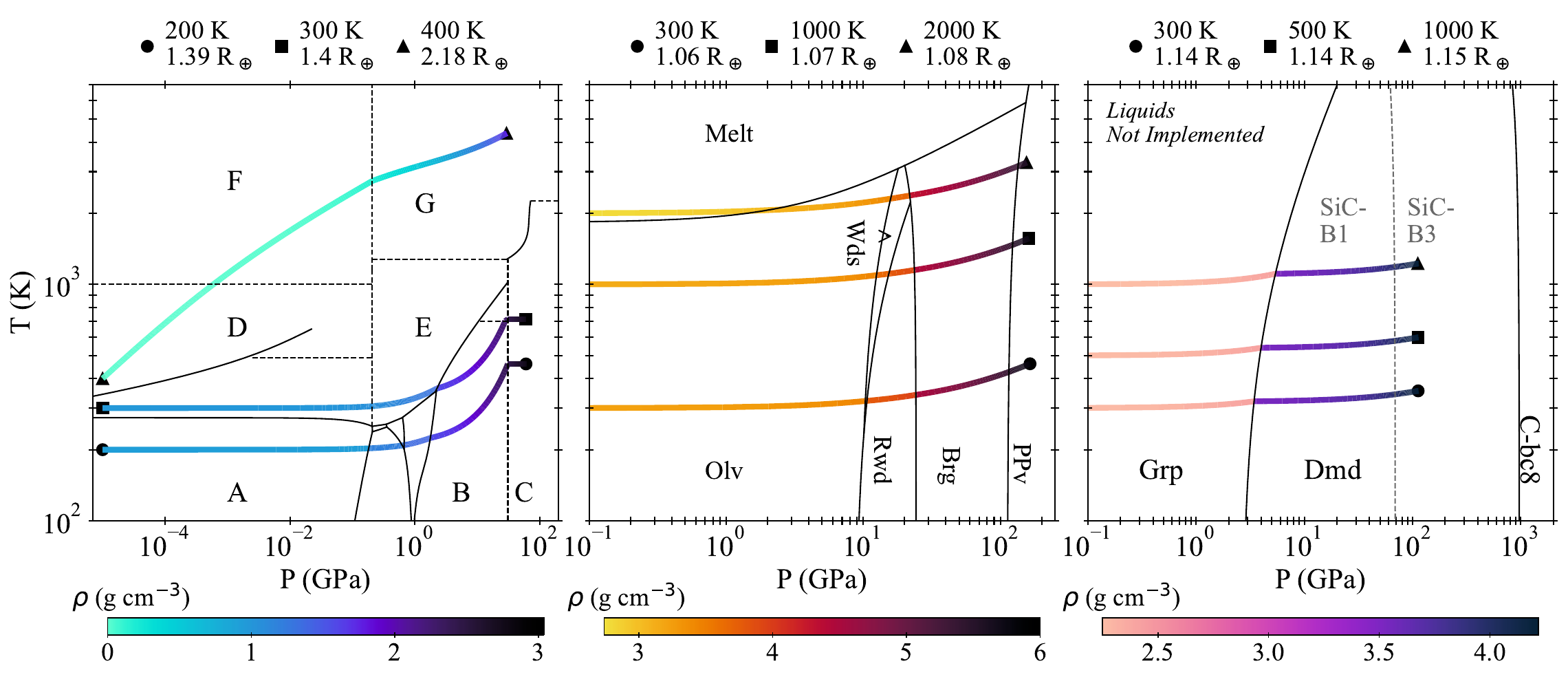}
\caption{New phase diagrams in the code. Left, updated default
hydrosphere. Center, default mantle with lower-pressure
Mg\(_2\)SiO\(_4\) phases. Right, carbon and SiC mantle phase diagrams.
On each plot are shown the pressure--temperature conditions inside a one
Earth-mass planet with different outer temperatures. The plotting
script, model output files, and instructions used to reproduce this
figure are available in the
\href{https://github.com/DavidRRice/MagratheaRelatedFiles/tree/main/JOSSPhaseDiagrams}{JOSSPhaseDiagrams
directory}. \label{fig:phases}}
\end{figure}

\begin{itemize}
\tightlist
\item
  \textbf{Default Mantle:} Added upper-mantle polymorphs of
  Mg\(_2\)SiO\(_4\) (forsterite, wadsleyite, ringwoodite)
  (\citeproc{ref-Dorogokupets:2015}{Dorogokupets et al., 2015}), see
  \autoref{fig:phases}.
\item
  \textbf{Default Hydrosphere:} Updated H\(_2\)O EOS and phase
  boundaries for ices (\citeproc{ref-Journaux:2020}{Journaux et al.,
  2020}), liquid and gas (\citeproc{ref-Wagner:2002}{Wagner \& Pruß,
  2002}), and supercritical water (\citeproc{ref-Mazevet:2019}{Mazevet
  et al., 2019}), largely inspired by the AQUA package
  (\citeproc{ref-Haldemann:2020}{Haldemann et al., 2020}), see
  \autoref{fig:phases}.
\item
  \textbf{Additional Gas EOS:} Including the solar-metallicity
  hydrogen/helium table from Chabrier \& Debras
  (\citeproc{ref-Chabrier:2021}{2021}) and van der Waals gases.
\item
  \textbf{Carbon Mantles:} EOS and phase diagrams for phases of carbon
  (\citeproc{ref-Benedict:2014}{Benedict et al., 2014};
  \citeproc{ref-Lowitzer:2006}{Lowitzer et al., 2006}) and silicon
  carbide (\citeproc{ref-Miozzi:2018}{Miozzi et al., 2018}), see
  \autoref{fig:phases}.
\item
  \textbf{EOS library growth:} Including the AQUA table
  (\citeproc{ref-Haldemann:2020}{Haldemann et al., 2020}), fcc- and
  bcc-iron (\citeproc{ref-Dorogokupets:2017}{Dorogokupets et al.,
  2017}), and mantle materials from Stixrude \& Lithgow-Bertelloni
  (\citeproc{ref-Stixrude:2011}{2011}).
\end{itemize}

\textbf{New functionality and solvers}

\begin{itemize}
\tightlist
\item
  \textbf{Composition finders:}

  \begin{itemize}
  \tightlist
  \item
    A secant-method routine that determines the mass of a third unknown
    layer given a target mass, radius, and ratio between the other two
    layers, looped over layer ratios and mass--radius posterior draws.
  \item
    A Markov chain Monte Carlo routine following Rogers \& Seager
    (\citeproc{ref-Rogers:2010}{2010}) and Dorn et al.
    (\citeproc{ref-Dorn:2015}{2015}) for probabilistic composition
    inference given mass, radius, and associated uncertainties with the
    Metropolis--Hastings method.
  \end{itemize}
\item
  \textbf{Tabulated EOS:} Support for tabulated
  \(P\)--\(T\)--\(\rho\)--\(\nabla T_S\) EOS tables using bilinear
  interpolation.
\item
  \textbf{Modular phase diagrams:} Users can store multiple
  phase-diagram configurations and call them in the configuration file,
  for example switching between silicate and carbon mantle models
  without recompiling.
\end{itemize}

\textbf{Usability}

\begin{itemize}
\tightlist
\item
  \textbf{Input handling:} All input parameters were moved to
  \texttt{run/*.cfg} files with descriptive keys.
\item
  \textbf{Parallelization:} Bulk runs and composition finder routines
  can exploit OpenMP in \texttt{compfind.cpp} enabling execution with
  multiple threads.
\item
  \textbf{Built-in numerical tests:} The \texttt{./planet\ -\/-test}
  option verifies pressure--density inversion for an analytical Vinet
  EOS and compares the structure solver for a constant-density planet
  against its analytical radius.
\item
  \textbf{Diagnostics:} More informative error messages are returned
  when solutions fail to converge.
\item
  \textbf{Tutorial and documentation:} A guided set of examples and
  practice problems resides in the \texttt{docs/} folder with online
  documentation at
  \href{https://magrathea.readthedocs.io}{magrathea.readthedocs.io}.
\end{itemize}

Together, these changes make Magrathea v2 a substantially expanded
platform rather than a simple update to the default planet model.

\section{Research impact statement}\label{research-impact-statement}

Magrathea has been used in published work to generate mass--radius
diagrams (\citeproc{ref-Taylor:2025}{Taylor et al., 2025}), infer
interiors of observed planets (\citeproc{ref-Desai:2024}{Desai et al.,
2024}; \citeproc{ref-MacDonald:2022}{MacDonald et al., 2022}), connect
theoretical composition from formation to observables
(\citeproc{ref-Childs:2023}{Childs et al., 2023};
\citeproc{ref-Dou:2024}{Dou et al., 2024};
\citeproc{ref-Steffen:2025}{Steffen et al., 2025}), and test how updated
high-pressure EOS measurements and interior assumptions change inferred
structures (\citeproc{ref-Huang:2021}{Huang et al., 2021};
\citeproc{ref-Lozovsky:2026}{Lozovsky, 2026}). The new composition
finders are used and described in Daspute et al.
(\citeproc{ref-Daspute:2025}{2025}), Rice et al.
(\citeproc{ref-Rice:2025}{2025}), and Kroft et al.
(\citeproc{ref-Kroft:2026}{2026}).

Beyond its use in published research, Magrathea is already contributing
to emerging intercomparison efforts in exoplanet interior modeling.
Schulze et al. (\citeproc{ref-Schulze:2026}{2026}) compares EOS and
material choices across common rocky-planet models and shows that those
choices can change inferred compositions at a level comparable to
current observational uncertainties. In that context, open and modular
tools like Magrathea help make differences in physical assumptions, EOS
choices, and retrieval workflows easier to isolate and test.

Magrathea now covers a broader set of interior assumptions, supports
faster composition inference workflows, and makes EOS-level sensitivity
studies easier to perform. These are the main research impacts of this
version.

\section{AI usage disclosure}\label{ai-usage-disclosure}

Generative AI was not used to implement most of the updates described in
this paper. Limited use of Generative AI was explored during development
of the MCMC mode. All AI-assisted code changes were reviewed by the
authors against the existing code base, and outputs were tested against
other non-AI retrieval modes. Generative AI was also used in drafting
and reorganizing this manuscript. All scientific claims and software
descriptions were checked and revised by the authors against prior
publications and the broader literature.

\section{Acknowledgements}\label{acknowledgements}

We acknowledge the many researchers using the Magrathea code base; we
appreciate each contribution to science and the code. We thank Douglas
Adams for the timeless stories compiled in ``The Hitchhiker's Guide to
the Galaxy'' which inspired the name of the code. We acknowledge support
from the College of Sciences and the Nevada Center for Astrophysics at
the University of Nevada, Las Vegas. A.V. acknowledges support by ISF
grants 770/21 and 773/21. C.H. is sponsored by Shanghai Pujiang Program
(grant No.~23PJ1414900).

\section*{References}\label{references}
\addcontentsline{toc}{section}{References}

\protect\phantomsection\label{refs}

\end{document}